%% file: main.tex
\PassOptionsToPackage{table}{xcolor}
\documentclass[letterpaper,twocolumn,10pt]{article}
\usepackage{usenix}
\usepackage{xurl}       
\usepackage{orcidlink}  

\usepackage{amsmath,amssymb}
\usepackage{graphicx}
\usepackage{booktabs}
\usepackage{multirow}
\usepackage[caption=false]{subfig}
\usepackage{array}
\usepackage{textcomp}
\usepackage{xspace}
\usepackage{tcolorbox}
\usepackage{wasysym}    
\usepackage{pifont}     
\usepackage{placeins}   
\usepackage{float}      
\usepackage{stfloats}   

\tcbuselibrary{breakable}   
\tcbset{top=3pt, bottom=3pt, left=5pt, right=5pt, before skip=6pt, after skip=6pt}

\newcommand{\etal}{\emph{et~al.}}

\graphicspath{{figures/}}

\begin{document}

\date{}
\title{\Large \bf Et Tu, MacBook? Unprivileged Keystroke Inference and\\Context Profiling via the Built-in IMU Side Channel}

\author{%
  {\rm Jiaji He}\textsuperscript{1}\orcidlink{0000-0003-1443-9279},
  {\rm Yi Shi}\textsuperscript{1}\orcidlink{0009-0006-8437-2742},
  {\rm Junfeng Cai}\textsuperscript{1}\orcidlink{0009-0004-4110-7151},\\[1pt]
  {\rm Chang Liu}\textsuperscript{2}\orcidlink{0000-0002-1834-4958},
  {\rm Yongqiang Lyu}\textsuperscript{1,3}\orcidlink{0000-0003-2573-963X}\\[3pt]
  \small\textsuperscript{1}Tianjin University \quad
  \textsuperscript{2}National University of Singapore \quad
  \textsuperscript{3}Tsinghua University\\[2pt]
  \small\textsuperscript{1}\{dochejj, shiyi2498, 3020232023\}@tju.edu.cn \quad
  \textsuperscript{2}fjmxlc@gmail.com \quad
  \textsuperscript{3}luyq@tsinghua.edu.cn%
}

\maketitle

\newcommand{\attack}{\textit{BRUTUS}\xspace}
\newcounter{kfcount}
\setcounter{kfcount}{1}
\newcommand{\kf}[1]{%
  \textbf{Security Insight \thekfcount: #1}\stepcounter{kfcount}\ignorespaces%
}

\input{sections/sec_abstract}

\input{sections/sec_introduction}
\input{sections/sec_background}
\input{sections/sec_threat_model}
\input{sections/sec_characterization}
\input{sections/sec_brutus}
\input{sections/sec_discussion}
\input{sections/sec_related}
\input{sections/sec_conclusion}

\appendix
\input{sections/sec_ethics}
\input{sections/sec_open_science}

\bibliographystyle{plainurl}
\bibliography{references}

\end{document}

%% file: sections/sec_abstract.tex
\begin{abstract}
Recent generations of Apple MacBooks embed an inertial measurement unit (IMU) within their unibody chassis for device orientation and motion sensing. However, this IMU inadvertently captures not only intended device-level information but also subtle physical vibrations from user interactions and the surrounding environment. These signals establish a novel, previously unexplored side channel. We uncover a vulnerability allowing non-root access to IMU data via an IOKit driver, alongside two content-free system metadata interfaces (HIDIdleTime and CGEventSource) that further enrich the side-channel leakage. Through rigorous characterization of the IMU data, we reveal that the leakage spans three core dimensions: (1) keystroke identity (which key is typed), (2) desk surface (where the laptop is placed), and (3) user behavior (who is typing). Leveraging these findings, we introduce \textit{BRUTUS}, the first comprehensive unprivileged side-channel attack targeting built-in IMU sensors on Apple MacBooks. \textit{BRUTUS} achieves a character-level accuracy of 89.1\% to 97.5\% in key recovery. Furthermore, aided by language models, it can successfully reconstruct certain sentences with 100\% accuracy. For user identification and environment profiling, \textit{BRUTUS} correctly discovers user and environment profiles without labels and correctly assigns subsequent segments to their corresponding profiles. Ultimately, this work highlights the urgent necessity of strictly regulating access to built-in IMU sensors.
\end{abstract}

%% file: sections/sec_introduction.tex
\section{Introduction}
\label{sec:introduction}

Mac devices equipped with Apple Silicon dominate the global personal computing and enterprise markets due to their outstanding performance and energy efficiency. Their massive market share and highly unified, closed-loop hardware design make them attractive targets for system security research. Consequently, the security community actively explores vulnerabilities in Apple Silicon, proposing numerous side-channel attacks (SCAs) against these platforms~\cite{kim2023ileakage,zhang2026tide,liu2026ssbench,harrison23,giallanza19,taneja2023hotpixels,hetterich2022branch_different,yu2023synchronization,kim2025slap,kim2025flop,ravichandran2022pacman,vicarte2022augury,chen2024gofetch}.

As the primary MacBook input modality, keyboard input carries sensitive content and is a longstanding side-channel target. Physical side-channel attacks (PSCAs) analyze keystroke-induced sound, electromagnetic or wireless effects, and vibration. They use external receivers or co-located sensors~\cite{asonov04,vuagnoux09,ali15,fang23,marquardt11,liu15,maiti16}, or access target-integrated microphones, cameras, accelerometers, and gyroscopes to recover keys, PINs, and text~\cite{harrison23,optivibe26,cai11,xu12taplogger,owusu12,miluzzo12,mehrnezhad17,ping15textlogger}. Software side-channel attacks (SSCAs) instead infer when keys are pressed, which keys are pressed, or what was typed from encrypted traffic, inter-keystroke intervals, and cache accesses~\cite{monaco19,song01,qiu26keytar,gruss15,schwarzl23lbta}.

Applying these attacks to Apple Silicon Macs faces practical obstacles. External physical observation requires additional hardware within sensing range. Target-integrated sensors avoid that deployment, but macOS protects camera and microphone access through TCC~\cite{apple-tcc} and exposes no supported API for third-party applications to read raw chassis accelerometer or gyroscope data~\cite{apple-coremotion,olvvier-medium}. SSCAs need no sensor hardware but depend on available software traces: traffic attacks require distinguishable input-driven communication~\cite{monaco19}; timing attacks expose intervals rather than key identities and require behavioral data and candidate or language priors~\cite{song01,qiu26keytar}; direct-key cache attacks require key-dependent accesses, version-specific profiling, and processor-specific probes~\cite{gruss15,schwarzl23lbta,yu2023synchronization,kim2023ileakage}. These constraints motivate examining MacBook-embedded sensors and their software access paths.

Recently, Bourbonnais et al.~\cite{olvvier-medium} discovered that Apple embeds an undocumented inertial measurement unit (IMU) in MacBooks to monitor device motion. They showed that a root-privileged process can access the raw IMU stream and use it to infer fine-grained physiological signals, including a user's pulse~\cite{olvvier-medium,olvvier-github}. Their implementation treats root privilege as a prerequisite for accessing the sensor data~\cite{olvvier-github}.

This discovery raises two immediate security questions. \textit{First, does the undocumented MacBook IMU expose an access path to unprivileged applications? Second, if so, what sensitive information can the raw sensor stream reveal?}

We find that prior work's assessment of the access boundary is overly optimistic~\cite{olvvier-github}: an unprivileged application can read raw IMU data through IOKit without root or runtime elevation. Section~\ref{sec:characterization} then shows that key position, strike force, and supporting surface leave distinguishable vibration features. Their persistence across MacBooks and evaluated noise conditions provides the basis for inferring input, user behavior, and usage context.

Based on these findings, we present \textbf{\attack} (\underline{B}uilt-in senso\underline{R} exploitation of \underline{U}ser \underline{T}yping via im\underline{U} \underline{S}ide channel), an unprivileged side-channel attack framework that uses keystroke-induced chassis vibrations captured by the undocumented MacBook IMU to infer typed content, typist-dependent characteristics, and the laptop's placement surface. We evaluate these three tasks in Sections~\ref{sec:brutus-keystroke}--\ref{sec:brutus-env}. On 8-to-10-character passwords from three held-out participants using held-out devices, \attack achieves an average character-level accuracy of 94.0\% and a Top-5 accuracy of 86.7\%; without user or environment labels, it also discriminates among typists, distinguishes among laptop placement surfaces, and assigns subsequent segments to their corresponding profiles.

\noindent \textbf{Contributions.}~
We summarize our contributions as follows:

\begin{itemize}
    \item We show that unprivileged applications can directly read raw data from the undocumented IMU built into MacBooks through IOKit without root privileges, runtime privilege elevation, or explicit user authorization, demonstrating that platform sensor access controls must also cover undocumented, vendor-internal sensor interfaces.

    \item We systematically characterize physical leakage from the MacBook IMU, demonstrating reliable distinctions among key positions, strike forces, and supporting surfaces and evaluating the effects of device variation and common noise. Based on these findings, we design \attack, which, to our knowledge, is the first side-channel attack framework to use an unprivileged local access path to the undocumented MacBook IMU for keystroke and context inference.

    \item We evaluate \attack on three tasks: keystroke inference, user profiling, and environment profiling. Our results show that \attack recovers typed content from held-out participants on unseen devices, identifies individual typists, and discriminates among laptop placement surfaces.
\end{itemize}

\noindent \textbf{Vulnerability Disclosure.}~
We reported the IMU vulnerability and the SCA vectors to Apple in March 2026. In their response in May 2026, \textbf{\textit{Apple acknowledged the vulnerability and confirmed the successful reproduction of the IMU data leakage}}. Apple is currently analyzing the root cause and developing a security patch to mitigate this issue.

%% file: sections/sec_background.tex
\section{Background}
\label{sec:background}

\subsection{Side-Channel Attacks}
\label{sec:bg-sca}

Side-channel attacks infer sensitive information from unintended signals accompanying system execution, device operation, or user activity. They have recovered cryptographic keys and memory contents, identified websites and application activity, inferred address layouts, and analyzed user interactions such as keystrokes~\cite{asonov04,taneja2023hotpixels,kim2023ileakage,jang2024sysbumps,chen2024gofetch}.

Physical side channels observe power consumption, electromagnetic and acoustic emissions, temperature, optical changes, motion, and mechanical vibration. Attackers capture these signals using dedicated instruments, nearby mobile or wearable devices, or software-readable sensors and telemetry on the target itself~\cite{asonov04,vuagnoux09,backes09,marquardt11,masti2015thermal}.

Software side channels exploit observable network, execution, or shared-resource states, including encrypted-traffic patterns, execution latency, caches, branch predictors, prefetchers, and interrupts~\cite{gruss15,hetterich2022branch_different,vicarte2022augury,kim2023ileakage,zhang2026tide}. They enable website and application fingerprinting, memory disclosure, address-layout inference, and cryptographic-secret recovery~\cite{taneja2023hotpixels,kim2023ileakage,jang2024sysbumps,chen2024gofetch}. Physical leakage and software observation may also be combined by reading sensors or telemetry through software interfaces or translating physical effects into software-observable timing~\cite{lipp2021platypus,wang2022hertzbleed,taneja2023hotpixels,miluzzo12}.

Practical deployment depends on signal observability and stability. Physical observations are affected by sensor placement, distance, line of sight, environmental noise, and attenuation; target-integrated sensors are further constrained by interface availability, access control, sampling rate, and background execution. Software channels depend on stable changes in communication, execution, or shared resources and may be constrained by network protocols, application implementations, binary layouts, processor architectures, and available timing or probing primitives.

\subsection{Inertial Measurement Units in MacBooks}
\label{sec:bg-imu}

An inertial measurement unit (IMU) typically combines a three-axis accelerometer with a three-axis gyroscope. The accelerometer measures linear acceleration along three spatial axes, whereas the gyroscope measures angular velocity about those axes, together providing six-axis inertial measurements. When mounted inside a laptop chassis, an IMU senses not only rigid-body translation and rotation but also mechanical vibrations transmitted through the structure.

Apple introduced the Sudden Motion Sensor (SMS) in its laptops in 2005. The SMS used a three-axis accelerometer to detect drops and protect mechanical hard drives~\cite{apple-sms}. Apple later phased out the SMS as MacBooks transitioned to solid-state storage. A subsequent logic-board teardown of the MacBook Air~M2 identified a Bosch Sensortec six-axis MEMS accelerometer and gyroscope~\cite{ifixit-m2-chipid}. Apple has not publicly documented the full role of this IMU in modern MacBooks, and macOS does not expose raw acceleration or angular-velocity data from the MacBook chassis through a supported third-party motion API~\cite{apple-coremotion,olvvier-medium}.

In 2026, Bourbonnais documented an IOKit-based method for accessing this IMU. The work identified the undocumented \texttt{AppleSPUHIDDevice} HID service and decoded its report format; the accompanying implementation runs as root and uses the resulting motion measurements to analyze fine-grained physiological signals such as pulse~\cite{olvvier-github,olvvier-medium}. \texttt{AppleSPUHIDDevice} thus provides a user-space data path from the MacBook's built-in IMU.

\begin{figure}[bt]
    \centering
    \includegraphics[width=\columnwidth]{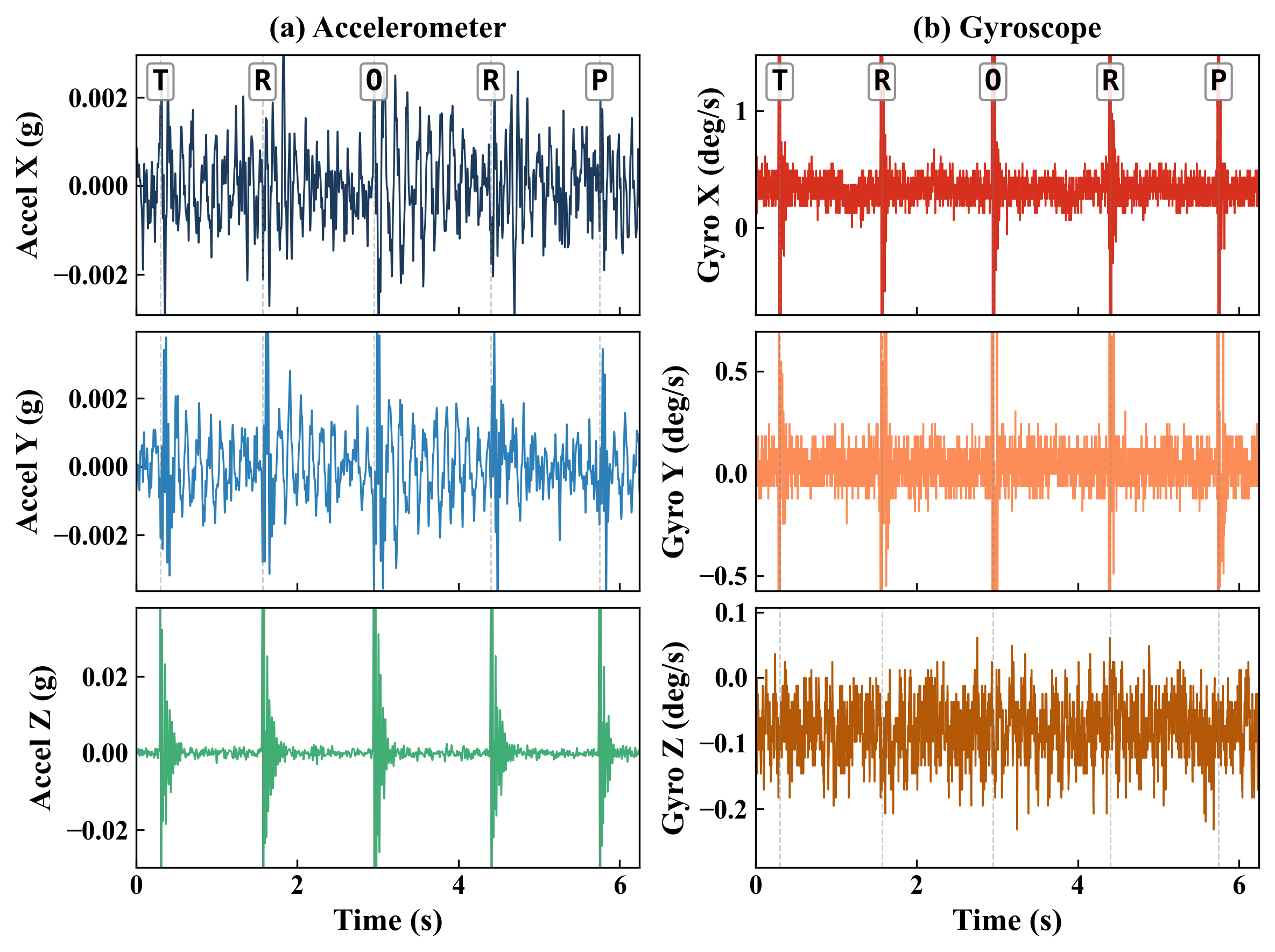}
    \caption{Raw six-axis IMU signal during password entry. Dashed lines mark keystroke onsets (keys T, R, O, R, P): (a)~accelerometer and (b)~gyroscope.}
    \label{fig:raw-imu}
\end{figure}

Figure~\ref{fig:raw-imu} shows a representative six-axis IMU trace that we collected through this data path during password entry. Each keystroke produces a sharp vibration transient that is visible in both the accelerometer and gyroscope channels, and waveform shapes exhibit observable differences across key positions. These traces show that the MacBook IMU senses not only rigid-body motion but also fine-grained structural vibrations generated by physical keyboard input. Once such physical measurements are delivered to user space, whether an ordinary application can read them depends on the account, login-session, and application-authorization checks that macOS applies to the data path.

\subsection{macOS Account and Session Model}
\label{sec:bg-macos-access}

macOS distinguishes standard and administrator accounts. The first user created during setup is automatically an administrator, making administrator-group membership the default for the initial Mac user~\cite{apple-users-groups}. Applications launched normally from that account still execute under the user's identifier (UID) and merely inherit its group memberships. Root execution requires a separate runtime-elevation mechanism such as \texttt{sudo}; administrator membership permits requesting elevation but does not make a normally launched application root~\cite{apple-terminal-sudo,apple-root-user}.

The \emph{active console user} occupies the Mac's current graphical desktop. Apple's \texttt{SCDynamicStoreCopyConsoleUser} returns that user's name and UID while excluding sessions switched out through Fast User Switching~\cite{apple-console-user}. Applications launched in the desktop session run under this console UID. Console status follows from login and neither grants administrator or root privileges nor requires a separate authorization.

macOS separately governs application capabilities through TCC and code-signing entitlements. TCC Input Monitoring controls whether an application can monitor keyboard, mouse, and trackpad input across applications~\cite{apple-tcc}; an entitlement embedded in the code signature grants a designated system capability~\cite{apple-entitlements}. We use \emph{unprivileged application} for an application running under the active console user's non-root UID, even when that user belongs to the administrator group, without runtime elevation, a special entitlement for the target interface, or TCC Input Monitoring authorization.

%% file: sections/sec_threat_model.tex
\section{Threat Model}
\label{sec:threat_model}

In this paper, we assume an unprivileged attacker who exploits the MacBook's built-in IMU side channel. The attacker aims to extract sensitive information, such as exact keystrokes, user identities, and the laptop's external environment.

\noindent \textbf{Operating environment.}~
We assume the victim operates an Apple MacBook equipped with a built-in IMU, running a standard, unmodified macOS installation. The experiments in this study are conducted on the latest MacBook models and macOS versions available at the time of writing. Furthermore, we assume the victim is the device owner and is logged in as the active console user using the administrator account created during standard macOS setup. As discussed in Section~\ref{sec:background}, this administrator-group membership is a pre-existing property of the victim's account; the malicious process does not request administrator credentials or trigger an authentication prompt.

\noindent \textbf{IMU data capturing and exfiltration.}~
In line with the threat models of conventional software-based side-channel attacks~\cite{chen2024gofetch,jang2024sysbumps,zhang2026tide}, we assume the attacker can execute an unprivileged background process under the victim's active user account without triggering any macOS authentication prompts. This malicious process interfaces with the IOKit driver to continuously capture IMU data streams at a sampling rate of 800\,Hz. Finally, we assume the attacker can exfiltrate the acquired sensor data via covert channels~\cite{cabuk04covert,shah06jitterbug} to perform the necessary offline analysis remotely.

%% file: sections/sec_characterization.tex
\section{Characterization of IMU}
\label{sec:characterization}

\subsection{Experimental Setup}
\label{sec:char-setup}

We conduct the IMU characterization experiments on 10 MacBook Air laptops, as summarized in Table~\ref{tab:devices}. All devices are equipped with a built-in IMU sampled at approximately 800\,Hz. The evaluation spans three chip generations (M3, M4, M5), two screen sizes (13'' and 15''), and three macOS releases (Sonoma 14, Sequoia 15, and Tahoe 26). Each device is assigned a fixed identifier (\textit{D1} through \textit{D10}) used throughout the paper to reference individual machines.

\begin{table}[t]
\centering
\caption{Evaluation devices. Serial numbers show only the first three characters for privacy.}
\label{tab:devices}
\renewcommand{\arraystretch}{1.15}
\setlength{\tabcolsep}{3pt}
\small
\begin{tabular*}{\columnwidth}{@{\extracolsep{\fill}}l l l l l@{}}
\toprule
\textbf{Device} & \textbf{Model} & \textbf{Chip} & \textbf{macOS} & \textbf{Serial} \\
\midrule
D1  & MacBook Air 13'' & \;M4 & Tahoe 26.5   & CT6xxxxxxx \\
D2  & MacBook Air 13'' & \;M4 & Tahoe 26.3   & JWKxxxxxxx \\
D3  & MacBook Air 13'' & \;M4 & Sequoia 15.5 & CD6xxxxxxx \\
D4  & MacBook Air 13'' & \;M5 & Tahoe 26.3   & KKXxxxxxxx \\
D5  & MacBook Air 15'' & \;M5 & Tahoe 26.3   & D04xxxxxxx \\
D6  & MacBook Air 15'' & \;M3 & Sonoma 14.7  & M3Pxxxxxxx \\
D7  & MacBook Air 15'' & \;M3 & Sequoia 15.4 & LD7xxxxxxx \\
D8  & MacBook Air 13'' & \;M5 & Tahoe 26.4   & HLGxxxxxxx \\
D9  & MacBook Air 15'' & \;M4 & Tahoe 26.5   & K7Lxxxxxxx \\
D10 & MacBook Air 13'' & \;M3 & Tahoe 26.3   & HXVxxxxxxx \\
\bottomrule
\end{tabular*}
\end{table}

To systematically characterize the IMU side channel, participant P1 records all datasets in this section under controlled conditions on device D1 (unless otherwise noted). During recording, macOS Input Monitoring is temporarily enabled to obtain ground-truth key labels and per-keystroke timestamps; this authorization is absent in the attack scenario. Experiments use either controlled single-key repetitions or a fixed passage from \emph{Pride and Prejudice} covering all 26~letters and the space bar. Table~\ref{tab:char-datasets} summarizes the four characterization datasets. Full participant recruitment (ten participants, P1--P10) and attack-specific dataset details are described in Section~\ref{sec:brutus-overview}.

\begin{table}[b]
\centering
\caption{Characterization datasets (Section~\ref{sec:characterization}). \emph{Passage}: fixed English excerpt covering all 26 letters and space. \emph{All~9}: nine desk surfaces in Section~\ref{sec:char-desk}.}
\label{tab:char-datasets}
\renewcommand{\arraystretch}{1.15}
\setlength{\tabcolsep}{3pt}
\footnotesize
\begin{tabular*}{\columnwidth}{@{\extracolsep{\fill}}lclll@{}}
\toprule
\textbf{Dataset} & \textbf{Section} & \textbf{Device} & \textbf{Surface} & \textbf{Content} \\
\midrule
Key position       & \S4.3      & D1      & Wood          & 50/key    \\
Typing force       & \S4.4      & D1      & Wood          & 50/level  \\
Desk material      & \S4.5      & D1      & All 9         & Passage   \\
Cross-device       & \S4.6      & D1--D10 & Wood          & Passage   \\
\bottomrule
\end{tabular*}
\end{table}

\subsection{Access Control of IMU Data}
\label{sec:char-access}

Prior work on the MacBook IMU~\cite{olvvier-github} assumed that root privileges are required to read the sensor, which would severely limit the attack surface. If that assumption held, an attacker would need to first escalate privileges before exploiting the IMU, making a side-channel attack less practical. We therefore systematically audit the real privilege boundaries of all IMU-related interfaces on macOS.

As discussed in Section~\ref{sec:bg-macos-access}, the initial device-owner account created during standard macOS setup belongs to the administrator group, and its UID becomes the current console UID when the owner logs in to the graphical session. A process launched normally in this session therefore inherits both conditions without executing as root, invoking privilege elevation, or triggering an authentication or TCC prompt.

Beyond the privilege question, another challenge motivates our interface selection. Prior keystroke side-channel work typically segments individual keystrokes from the sensor stream using signal-level energy thresholding~\cite{asonov04, harrison23, radkey26}. However, the built-in IMU is rigidly coupled to the entire laptop chassis and records not only keystroke vibrations but also trackpad interactions, palm contacts, and other mechanical events at comparable amplitudes. In our preliminary experiments, the peak energy distributions of detected keyboard and trackpad events exhibited substantial overlap, and no energy threshold could reliably separate the two event types: at moderate sensitivity, a large fraction of detected events were trackpad interactions rather than keystrokes. Liu~\etal~\cite{liu15} observed a similar limitation on smartwatch accelerometers and resorted to a co-located microphone for segmentation. We therefore ask whether macOS exposes any content-free system interface that could provide precise keystroke timing from the software layer, bypassing the signal-level segmentation problem entirely.

macOS protects conventional input-event access through the TCC-protected \texttt{CGEventTap} and \texttt{IOHIDManager} interfaces, which expose raw key values and require explicit consent through the system-level Input Monitoring dialog~\cite{apple-tcc}.

We enumerate all IOKit HID services and CoreGraphics event interfaces that could leak keystroke-related information without revealing key values. Three interfaces survive this filter: \texttt{HIDIdleTime} provides sub-millisecond keystroke timestamps, \texttt{CGEventSourceSecondsSinceLastEventType} discriminates keyboard events from trackpad input, and \texttt{AppleSPUHIDDevice} delivers the raw six-axis IMU stream. Table~\ref{tab:access} contrasts these interfaces with the TCC-protected interfaces. The three interfaces used by BRUTUS reveal no key values and require neither root nor TCC authorization.

\begin{table}[t]
\centering
\caption{macOS input-related interfaces and their access conditions.}
\label{tab:access}
\renewcommand{\arraystretch}{1.15}
\setlength{\tabcolsep}{3pt}
\footnotesize
\begin{tabular*}{\columnwidth}{@{\extracolsep{\fill}}lccl@{}}
\toprule
\textbf{Interface} & \textbf{Key Val.} & \textbf{TCC} & \textbf{Access Cond.} \\
\midrule
\multicolumn{4}{@{}l}{\textit{TCC-protected}} \\
\texttt{CGEventTap}               & \ding{51} & \ding{51} & Console user \\
\texttt{IOHIDManager}             & \ding{51} & \ding{51} & Console user \\
\midrule
\multicolumn{4}{@{}l}{\textit{Exploited in this work}} \\
\texttt{HIDIdleTime}              & \ding{55} & \ding{55} & None \\
\texttt{CGEventSource}$^{a}$      & \ding{55} & \ding{55} & Console user \\
\texttt{AppleSPUHIDDevice}        & \ding{55} & \ding{55} & Admin group + console UID \\
\bottomrule
\end{tabular*}\\[2pt]
\raggedright\scriptsize
$^{a}$\,\texttt{SecondsSinceLastEventType}: keyboard/trackpad discrimination.
\end{table}

The three interfaces impose progressively stricter access conditions: unrestricted local access, an active GUI login session, and the combination of administrator-group membership and active-console-user status, respectively. First, \texttt{HIDIdleTime}, an IOKit Registry property on the \texttt{IOHIDSystem} service node, is readable by any local process via \texttt{IORegistryEntryCreateCFProperty}, regardless of privilege level, session context (GUI, SSH, or launchd), or TCC state. It resets to zero on every HID event; polling it at over 100\,kHz yields sub-millisecond keystroke timestamps.

Second, \texttt{CGEventSourceSecondsSinceLastEventType}, which discriminates keyboard events from trackpad input, requires only an active GUI login session. It imposes no administrator-group requirement and requires no TCC authorization.

Finally and most importantly, \texttt{AppleSPUHIDDevice} delivers the raw six-axis IMU stream at up to 800\,Hz. Its data path traverses four IOKit API stages. \texttt{IOServiceGetMatchingService} locates the \texttt{AppleSPUHIDDevice} node, and \texttt{IOHIDDeviceCreate} constructs a device reference. \texttt{IOHIDDeviceOpen} opens the device for reading and checks administrator-group membership. \texttt{IOHIDDeviceRegisterInputValueCallback} registers the data stream and performs a per-delivery User ID (UID) check that matches the caller against the current console owner. Our experiments confirm that these checks are enforced at runtime: an administrator-group process connected over SSH or running as a launchd daemon under a different account receives no callbacks, and a Fast User Switch suspends delivery to the original account within 3\,seconds.

We audited all three interfaces across the ten devices in Table~\ref{tab:devices}, spanning three Apple Silicon generations (M3, M4, M5) and three macOS releases (Sonoma, Sequoia, Tahoe); all findings are consistent across configurations. We also verified that toggling Input Monitoring authorization has no effect on the direct \texttt{AppleSPUHIDDevice} path, which continues to deliver approximately 800\,Hz accelerometer and gyroscope data in both states. This result shows that TCC Input Monitoring protects conventional input-event interfaces that expose key values, but does not cover the IMU data path exploited here.

\begin{tcolorbox}[colframe=black, boxrule=0.8pt, width=\linewidth, arc=1mm, auto outer arc, breakable]
\kf{} Contrary to prior assumptions~\cite{olvvier-github} that root privileges are required, a non-root process in the common device-owner session can access all three interfaces without TCC authorization, a dedicated entitlement, or any user-visible indicator, leaving this side-channel path outside the Input Monitoring gate that protects conventional input-event APIs.
\end{tcolorbox}

\subsection{Sensitivity to Key Positions}
\label{sec:char-key-position}

\begin{figure}[t]
    \centering
    \includegraphics[width=\columnwidth]{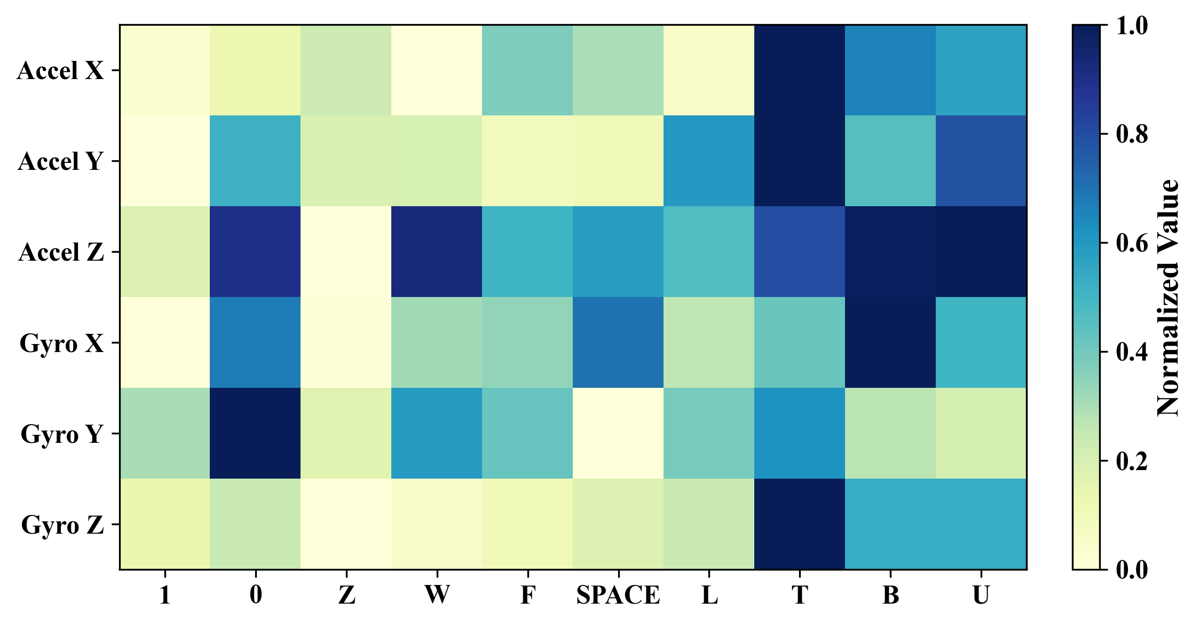}
    \caption{Normalized IMU response across six channels for ten keys.}
    \label{fig:key-position}
\end{figure}

Pressing a key transmits vibrations through the laptop chassis. Because the IMU is mounted off-center on the logic board~\cite{idownloadblog-m2}, each key has a different distance and angle to the sensor, creating a distinct mechanical lever arm. The resulting vibration pattern therefore differs from key to key in both amplitude and phase across the six IMU axes~\cite{marquardt11}. To quantify this, we select ten representative keys spanning the full keyboard layout (1, 0, Z, W, F, SPACE, L, T, B, U) and record 50 keystrokes per key on D1. To minimize interference from force variation, a single operator (P1) types all keystrokes within the same session at a deliberately steady pace. Figure~\ref{fig:key-position} visualizes the normalized standard deviation of each IMU channel per key. Every key exhibits a distinct six-channel signature, confirming that key position is reliably encoded in the IMU signal. This per-key discriminability forms the physical basis for keystroke inference.

\begin{tcolorbox}[colframe=black, boxrule=0.8pt, width=\linewidth, arc=1mm, auto outer arc, breakable]
\kf{} Different key positions produce distinguishable six-channel IMU signatures, providing the physical basis for keystroke inference from a single vibration window.
\end{tcolorbox}

\subsection{Sensitivity to Typing Forces}
\label{sec:char-typing-force}

\begin{figure}[t]
    \centering
    \includegraphics[width=\columnwidth]{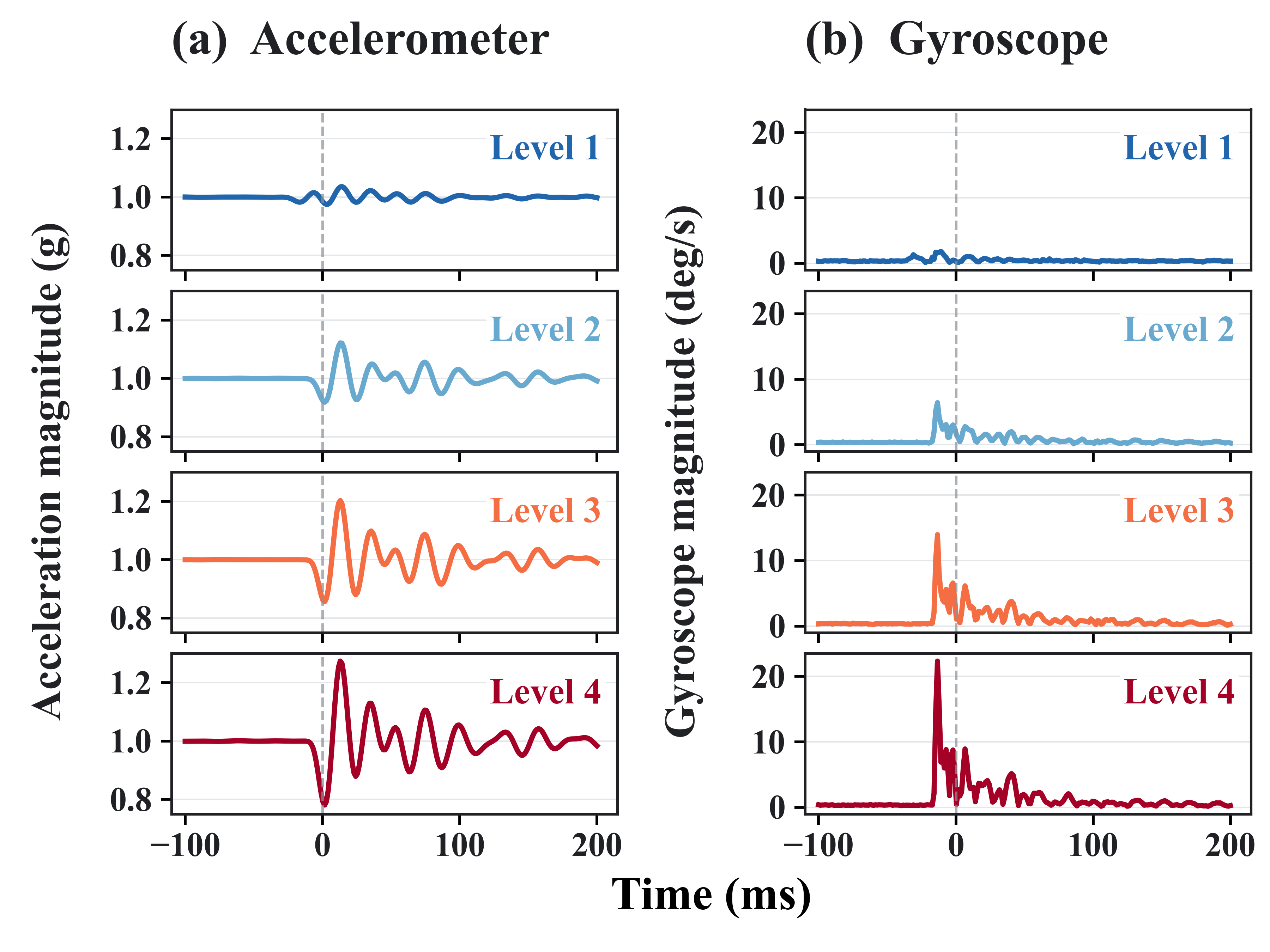}
    \caption{Waveforms of key M at four force levels. (a)~Accelerometer magnitude. (b)~Gyroscope magnitude.}
    \label{fig:typing-force}
\end{figure}

Typing force varies across keystrokes: a harder press increases the impulse and vibration amplitude, whereas the key mechanism and chassis determine the waveform's oscillation and decay. We test this separation by pressing key M 50 times at each of four force levels, from light touch to deliberate hard press, and compute orientation-independent accelerometer and gyroscope magnitudes:
\begin{equation}
\|\mathbf{a}\| = \sqrt{a_x^2 + a_y^2 + a_z^2}, \quad
\|\boldsymbol{\omega}\| = \sqrt{\omega_x^2 + \omega_y^2 + \omega_z^2}.
\label{eq:magnitude}
\end{equation}
Figure~\ref{fig:typing-force} shows that force scales amplitude while preserving the oscillation pattern, peak timing, and decay profile, allowing shape-based key features to generalize across natural force variation.

\begin{tcolorbox}[colframe=black, boxrule=0.8pt, width=\linewidth, arc=1mm, auto outer arc, breakable]
\kf{} Force variation scales only the vibration amplitude; the waveform shape that encodes key identity is preserved. A classifier trained on shape-based features can therefore generalize across natural force variation.
\end{tcolorbox}

\subsection{Impact of Desk Materials}
\label{sec:char-desk}

\begin{figure}[t]
    \centering
    \includegraphics[width=\columnwidth]{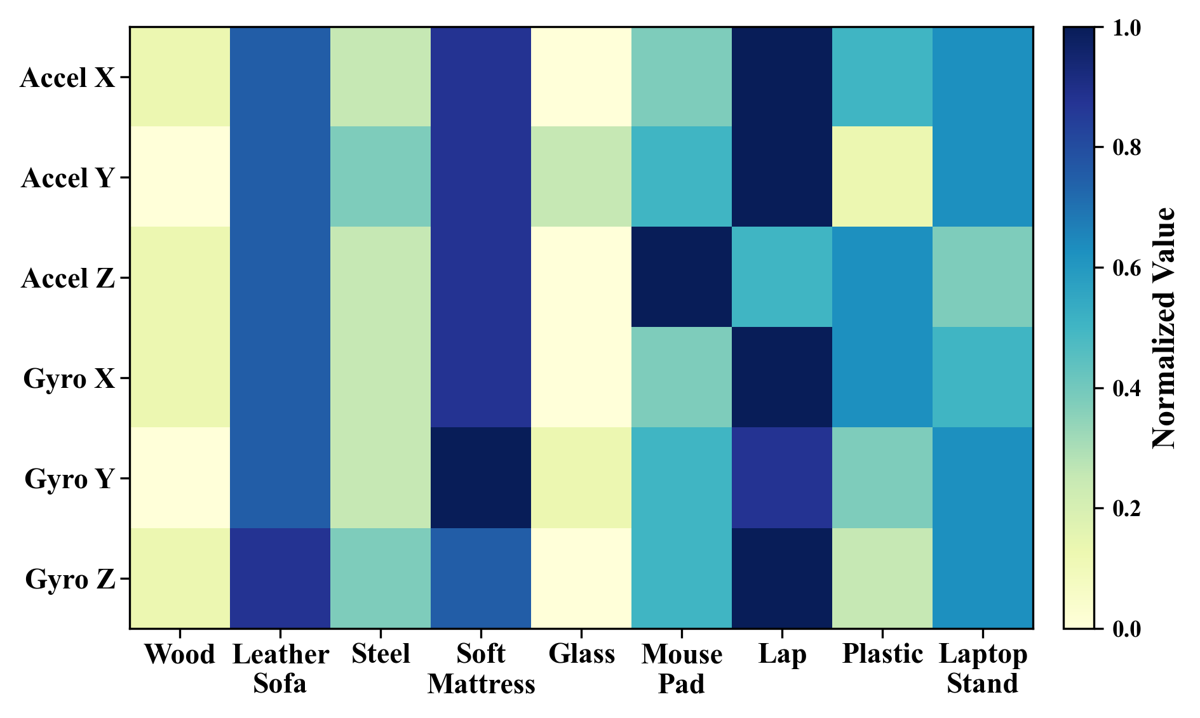}
    \caption{Normalized standard deviation of each IMU channel across nine desk surfaces.}
    \label{fig:desk-material}
\end{figure}

The surface beneath the laptop affects how keystroke vibrations propagate through the chassis. On a rigid surface (e.g., wood, steel), the laptop is mechanically well-coupled to the desk and keystroke energy dissipates rapidly into the supporting structure, producing lower residual vibration at the IMU. On a compliant surface (e.g., mattress, sofa, lap), the laptop rests on a yielding base that reflects rather than absorbs the impulse; the chassis oscillates longer and with greater amplitude~\cite{cremer2005structure}. We type the passage (Table~\ref{tab:char-datasets}) on D1 at consistent typing force across nine surfaces: four rigid (wood, glass, steel, plastic) and five non-rigid (leather sofa, soft mattress, mouse pad, lap, laptop stand with structural compliance). Figure~\ref{fig:desk-material} plots the normalized standard deviation of each IMU channel per surface. Consistent with the physics above, compliant surfaces (mattress, sofa, lap) exhibit higher normalized variability than rigid surfaces (wood, glass, steel), confirming the predicted coupling behavior. Each surface produces a visually distinct six-channel signature, confirming that the desk material leaves a measurable fingerprint in the IMU signal. This observation motivates environment-aware training for keystroke inference (Section~\ref{sec:brutus}) and enables surface identification as a standalone profiling capability.

\begin{tcolorbox}[colframe=black, boxrule=0.8pt, width=\linewidth, arc=1mm, auto outer arc, breakable]
\kf{} Each of nine desk surfaces produces a distinct six-channel IMU fingerprint. This enables environment classification as a standalone attack and necessitates multi-surface training for robust keystroke inference.
\end{tcolorbox}

\subsection{Cross-Device Consistency}
\label{sec:char-cross-device}

Cross-device attacks require the key-dependent signatures in Sections~\ref{sec:char-key-position}--\ref{sec:char-desk} to reflect the shared keyboard-chassis design rather than individual units.

P1 types the passage (Table~\ref{tab:char-datasets}) on D1--D10 on a wood desk, yielding approximately 900 keystrokes per device across 27 keys. The pipeline in Section~\ref{sec:brutus-overview} converts each keystroke into $\mathbf{F}_k \in \mathbb{R}^{240 \times 12}$, and the same InceptionTime backbone maps it to a 128-dimensional embedding. We train on D1--D7 with supervised contrastive learning, pulling same-key embeddings together across devices and separating different keys. Held-out D8--D10 span M3--M5 and both 13'' and 15'' chassis. Using cosine similarity, we report ROC AUC for pairwise same-key versus different-key discrimination.

\begin{figure}[t]
    \centering
    \includegraphics[width=\columnwidth]{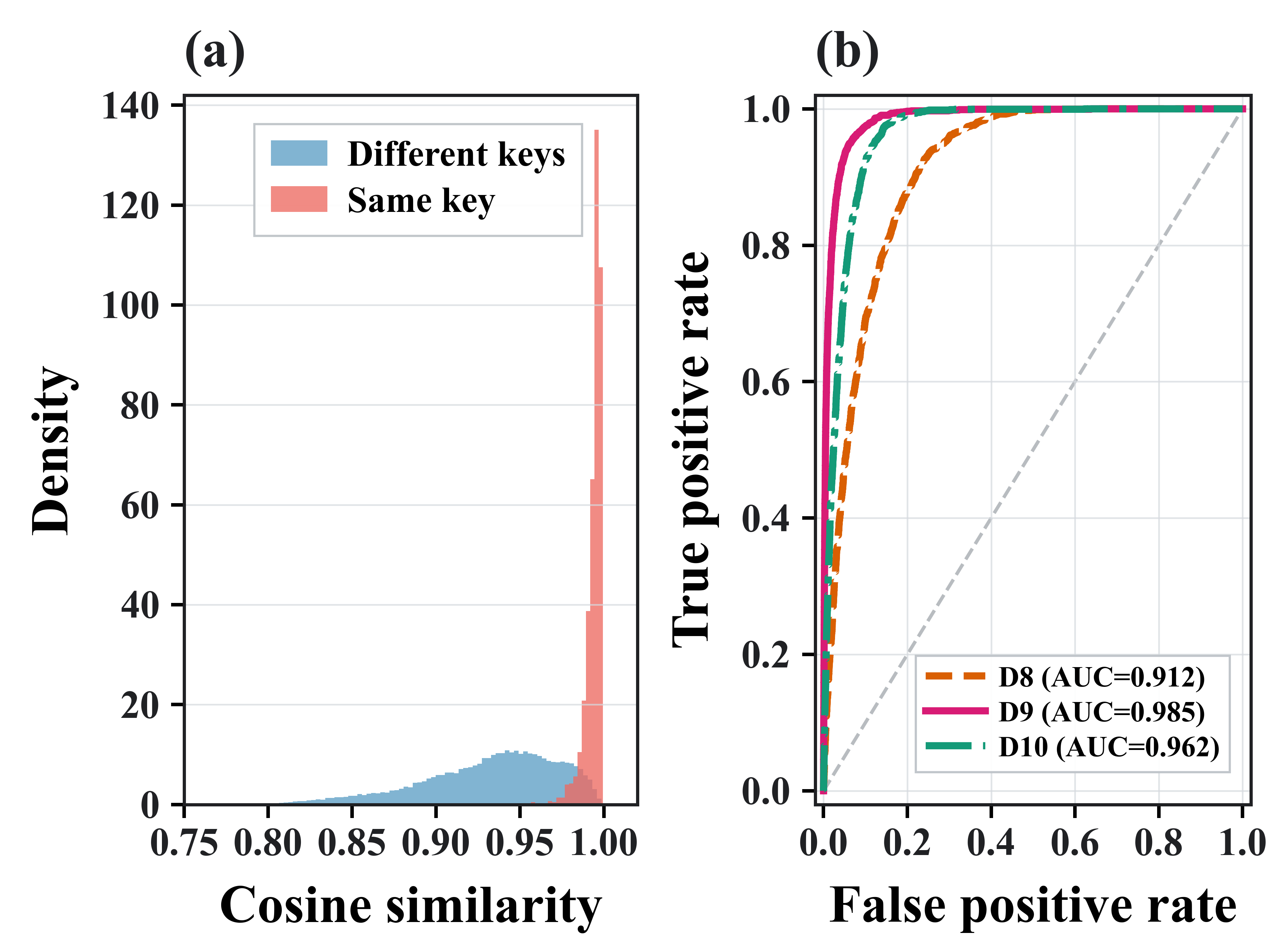}
    \caption{Cross-device keystroke discrimination on held-out devices. (a)~Cosine similarity distributions on D9: same-key pairs concentrate near 1.0, while different-key pairs spread below 0.95. (b)~ROC curves for pairwise discrimination on D8, D9, and D10.}
    \label{fig:cross-device}
\end{figure}

All held-out devices exceed 91\% AUC: D8, D9, and D10 achieve 91.2\%, 98.5\%, and 96.2\%, respectively. On D9, same-key similarities concentrate near 0.994 and different-key pairs near 0.932 (Figure~\ref{fig:cross-device}(a)); Figure~\ref{fig:cross-device}(b) shows consistent ROC separation across all three devices. The results cover three chip generations, two chassis sizes, and three macOS releases, supporting that key-dependent representations transfer across the tested MacBook designs rather than reflecting a single unit.

\subsection{Effects of Noise}
\label{sec:char-noise}

To assess robustness under common operating conditions, we used offline additive-noise injection: recorded noise was added to the test windows, after which pairwise AUC was recomputed~\cite{do2023nonprofiled}.

We recorded 120-second, keystroke-free IMU traces under four conditions: a phone on the same desk playing music at 50--60\% volume (External Speaker), the MacBook's built-in speakers at a comparable system volume (MacBook Speaker), dense local TCP traffic (Network Traffic), and four sustained SHA-256 workers (CPU Load). System-activity traces were recorded in the order idle, Network Traffic, and CPU Load, with quiet intervals of 30 and 60 seconds before the latter two recordings.

For each IMU channel, we estimated the contribution of the tested condition from the difference in AC power between the condition and idle recordings~\cite{boll1979suppression}:
\begin{equation}
    \sigma_{\mathrm{source}}
    = \sqrt{\max\!\left(\sigma_{\mathrm{condition}}^2
      - \sigma_{\mathrm{idle}}^2,\,0\right)}.
    \label{eq:noise-source-amplitude}
\end{equation}
We centered each condition trace and scaled it to the resulting source amplitude. Using one-second blocks, we performed block resampling~\cite{kunsch1989jackknife} and computed Bonferroni-adjusted one-sided simultaneous lower bounds across the six channels~\cite{dunn1961multiple}; a channel was injected only when its lower bound on variance increase over idle was positive.

For each test window, we added a 240-sample source segment to the six raw channels at $1\times$, $5\times$, $10\times$, or $25\times$ amplitude and recomputed the first-difference channels~\cite{yin2015noisy}. Each condition used five deterministic draws, with the same segments reused across amplification levels.

At the recorded amplitude ($1\times$), Table~\ref{tab:noise} shows mean AUC changes below 0.03 percentage points under every condition, within variation across the five draws. External Speaker, Network Traffic, and CPU Load remained close to the clean result even at $25\times$, with a maximum decrease of 0.25 percentage points. Only MacBook Speaker caused clear degradation, lowering AUC to 95.0\%, 76.9\%, and 52.1\% at $5\times$, $10\times$, and $25\times$. Its vibrations originate inside the chassis and share the mechanical path that carries keystroke vibrations. Pairwise discrimination therefore remained stable under the recorded conditions; substantial degradation occurred only for amplified, chassis-coupled speaker vibration.

\begin{table}[t]
\centering
\caption{Pairwise AUC (\%) under offline noise injection, averaged over five deterministic draws.}
\label{tab:noise}
\begin{tabular*}{\columnwidth}{@{\extracolsep{\fill}}lcccc@{}}
\toprule
Condition & $1\times$ & $5\times$ & $10\times$ & $25\times$ \\
\midrule
None (clean)    & 98.7 & 98.7 & 98.7 & 98.7 \\
External Speaker & 98.7 & 98.7 & 98.7 & 98.7 \\
MacBook Speaker  & 98.7 & 95.0 & 76.9 & 52.1 \\
Network Traffic  & 98.7 & 98.7 & 98.7 & 98.5 \\
CPU Load          & 98.7 & 98.7 & 98.7 & 98.7 \\
\bottomrule
\end{tabular*}
\end{table}

%% file: sections/sec_brutus.tex
\section{\attack: IMU Side-channel Attacks}
\label{sec:brutus}

Building on Section~\ref{sec:characterization}, \attack infers \emph{what} is typed (Section~\ref{sec:brutus-keystroke}), \emph{who} is typing (Section~\ref{sec:brutus-user}), and \emph{where} the laptop is placed (Section~\ref{sec:brutus-env}). A timing oracle segments the shared six-axis IMU stream. Keystroke inference classifies individual windows, whereas label-free profiling aggregates inter-channel coupling over typing segments to discover recurring groups.

\subsection{Attack Overview}
\label{sec:brutus-overview}

\noindent \textbf{Participants and datasets.}
We recruit ten regular laptop users (P1--P10), aged 22--35 with an equal sex split. P1 performs the characterization (Section~\ref{sec:characterization}) and environment profiling; P1--P8 contribute to user profiling, and P1--P7 provide keystroke-inference training data. P8--P10 contribute no such training data and serve as held-out victims (Section~\ref{sec:brutus-keystroke-eval}). Each training session adds 20 random passwords (10 each of lengths~8 and~9) to the passage, covering 37 classes and approximately 1{,}070 keystrokes. Desk-material characterization and environment profiling share recordings.


\noindent \textbf{Feature representation.}
A physical keystroke generates a transient mechanical impulse in the laptop chassis, producing a sharp onset, rapid resonant peak, and damped decay. Section~\ref{sec:char-typing-force} shows that strike force scales the amplitude without changing this waveform shape, which returns to the noise floor within approximately 200\,ms.

We therefore extract a 300\,ms asymmetric window around each onset timestamp $t_k$, obtained from labeled training data or the real-time timing oracle in Section~\ref{sec:brutus-keystroke}. Sampling $[t_k - 100\,\text{ms},\; t_k + 200\,\text{ms}]$ from the continuous approximately 800\,Hz IMU stream yields $\mathbf{W}_k \in \mathbb{R}^{240 \times 6}$, comprising three accelerometer and three gyroscope axes. The 100\,ms pre-onset interval provides a quiescent baseline, while the 200\,ms post-onset interval covers the impact and decay.

To encode the dynamics of each vibration channel beyond raw amplitude, we augment each window with first-order temporal differences:
\begin{equation}
  \mathbf{D}_k[t,c] = \mathbf{W}_k[t{+}1,c] - \mathbf{W}_k[t,c]
  \label{eq:temporal-diff}
\end{equation}
The first-order differences capture the key-dependent onset slopes and decay rates identified in Section~\ref{sec:char-key-position} and complement the raw amplitude profiles. Per-channel z-score normalization, using $\mu$ and $\sigma$ computed on the training data and reused at inference, yields $\mathbf{F}_k \in \mathbb{R}^{240 \times 12}$ for the keystroke classifier and supervised characterization models. The label-free attacks instead derive the segment representation in Section~\ref{sec:brutus-user} directly from the six raw channels.

\noindent \textbf{Classifier architecture.}
Under the same cross-validation protocol, we compared InceptionTime~\cite{ismail20}, a three-layer 1D CNN, a Transformer encoder, and a gradient-boosted ensemble on a held-out keystroke dataset. InceptionTime achieved 4--14 percentage points higher top-1 accuracy than the alternatives while using the fewest trainable parameters among the neural candidates, so we use it for keystroke inference.

InceptionTime extends the Inception module~\cite{inception15} to one-dimensional time series using parallel convolutions at multiple temporal scales. Our configuration has three cascaded Inception blocks with kernels of 10, 20, and 40 samples (approximately 12.5\,ms, 25\,ms, and 50\,ms at 800\,Hz), a bottleneck dimension of 32, and a max-pooling branch. The shorter kernels resolve impact transients, while the longer kernels capture chassis resonance and decay (Sections~\ref{sec:char-key-position} and~\ref{sec:char-desk}). Global average pooling produces a 128-dimensional vector, which a linear head maps to 37 key classes. The model has approximately 480K trainable parameters. Section~\ref{sec:char-cross-device} uses the same backbone with a supervised contrastive objective and confirms that it captures key-dependent structure across three chip generations.

\noindent \textbf{Training methodology.}
The supervised models use categorical cross-entropy and Adam (initial learning rate $10^{-3}$, weight decay $10^{-4}$), with cosine annealing and early stopping. We use five-fold, session-level GroupKFold cross-validation: all windows from a recording session remain in the same fold. The label-free attacks require no classifier training; their clustering procedure is specified in Section~\ref{sec:brutus-user}.

\begin{figure}[t]
    \centering
    \captionsetup{skip=0pt}
    \includegraphics[width=\columnwidth]{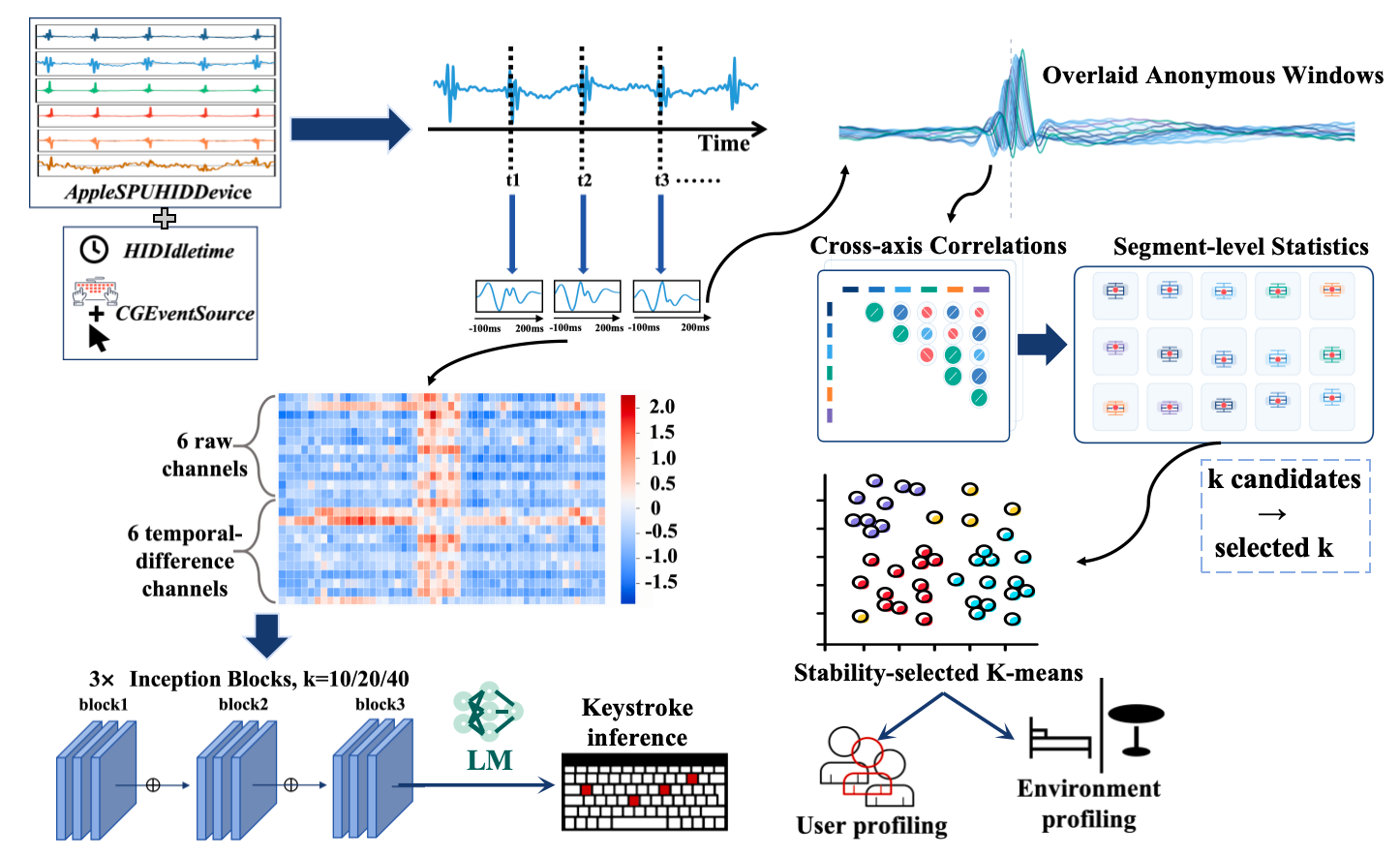}
    \par\vspace{-4pt}
    \caption{End-to-end \attack pipeline. The timing oracle segments the six-axis IMU stream into keystroke windows. Keystroke inference uses 12-channel features, InceptionTime, and language-model decoding; label-free profiling instead aggregates cross-axis correlations over typing segments, selects $k$ by prediction strength, and clusters recurring user or environment profiles with $k$-means.}
    \label{fig:attack-overview}
\end{figure}

\subsection{Keystroke Inference}
\label{sec:brutus-keystroke}

Keystroke inference recovers the exact character sequence typed by the victim. Against three held-out participants on unseen devices and in self-selected environments, \attack achieves an overall Character Error Rate (CER) of 2.5\%--10.9\% (Table~\ref{tab:e2e-keystroke}).

\noindent \textbf{Challenge and solution.}
\label{sec:brutus-insight}
Recovering a complete sequence first requires locating each keystroke in the continuous IMU stream. IMU-only segmentation is unreliable because successive vibrations overlap and ambient transients can mimic keystroke onsets.

\attack combines the content-free metadata interfaces characterized in Section~\ref{sec:char-access} to obtain precise onset timestamps without exposing key values. It polls \texttt{IOHIDSystem HIDIdleTime} at over 100{,}000 iterations per second and records each HID reset. A 4\,ms debounce filter merges the key-down/key-up resets from one physical press, after which \texttt{CGEventSourceSecondsSinceLastEventType} removes trackpad events. The resulting timestamps segment the continuous IMU stream into per-keystroke windows.

\noindent \textbf{Per-keystroke classification.}
For each retained event, the sub-millisecond timestamp indexes the IMU buffer to extract the 300\,ms, 12-channel window defined in Section~\ref{sec:brutus-overview}. InceptionTime classifies it over 37 keys (a--z, 0--9, space) and retains the full softmax distribution for sequence decoding.

\noindent \textbf{Sequence-level text recovery.}
The per-position softmax distributions are assembled into a ranked list of candidate strings via beam search with width $B{=}100$ and per-position expansion $K{=}6$: at each character position, the $K$ highest-probability classes extend the current $B$ partial sequences by cumulative log-probability, and only the top $B$ extensions survive to the next position. When the input contains linguistic structure, a trigram language model augments the search at word boundaries. The language model is trained on the Brown corpus ($\sim$1M tokens) with Laplace smoothing, and its vocabulary is supplemented by the NLTK English word list ($\sim$236K entries). The combined score at each word boundary balances classifier evidence and linguistic plausibility:
\begin{equation}
  S_\mathrm{word}(w) = S_\mathrm{cls}(w) + \alpha \cdot \log P_\mathrm{LM}(w \mid w_{-2}, w_{-1})
  \label{eq:word-score}
\end{equation}
where $S_\mathrm{cls}(w)$ is the cumulative character-level log-probability and $\alpha = 0.5$. A sensor-constrained rescue mechanism prevents the language model from overriding high-confidence classifier predictions: character positions where the softmax probability exceeds a threshold are frozen, and candidate words are accepted only if their per-position characters fall within the classifier's top-$K$ predictions at the majority of positions.

\noindent \textbf{Evaluation metrics.}
We report two metrics for keystroke inference. \emph{Character Error Rate} (CER) is the edit distance between the predicted and ground-truth character sequences, normalized by the ground-truth length; it captures substitutions, insertions, and deletions. \emph{Top-5 accuracy} is the fraction of test sequences for which the correct string appears among the five highest-scoring candidates produced by beam search.

\noindent \textbf{Results.}
\label{sec:brutus-keystroke-eval}
We train on the Section~\ref{sec:brutus-overview} data from P1--P7 operating D1--D7 across the surfaces in Section~\ref{sec:char-desk}.

We first evaluate generalization among the training participants: each of P1--P7 switches to a MacBook not used during their own training sessions and independently selects a typing environment, then types 15 random alphanumeric passwords (five each of length~8, 9, and~10) followed by ten meaningful English sentences. The upper portion of Table~\ref{tab:e2e-keystroke} reports the per-participant cross-test results.

To simulate a realistic attack against unseen victims, three participants (P8--P10) who did not contribute to the keystroke inference training set are each assigned a MacBook not used during training (D8--D10) and independently select a natural typing environment: P8 types at a caf\'{e} table, P9 on their lap, and P10 on a soft mattress. Each follows the same protocol: 15 passwords (five each of length~8, 9, and~10), then ten sentences. The lower portion of Table~\ref{tab:e2e-keystroke} reports these end-to-end results. Notably, P8 achieves an overall CER of 2.5\%, comparable to the best cross-test participants, demonstrating that the model does not degrade simply because the user, device, and environment are all unseen during training.

\begin{table}[t]
\centering
\caption{Keystroke inference results (\%). Overall combines password and sentence CER. Averages are weighted by character count.}
\label{tab:e2e-keystroke}
\renewcommand{\arraystretch}{1.15}
\setlength{\tabcolsep}{3pt}
\footnotesize
\begin{tabular*}{\columnwidth}{@{\extracolsep{\fill}}lll cc cc c@{}}
\toprule
& & & \multicolumn{2}{c}{\textbf{Password}} & \multicolumn{2}{c}{\textbf{Sentence}} & \\
\cmidrule(lr){4-5} \cmidrule(lr){6-7}
\textbf{User} & \textbf{Device} & \textbf{Surface} & \textbf{CER} & \textbf{Top-5} & \textbf{CER} & \textbf{+LM} & \textbf{Overall} \\
\midrule
\multicolumn{8}{@{}l}{\emph{Cross-test}} \\
P1  & D3  & Wood         & 7.4  & 100.0 & 5.5 & 0.8 & 5.8 \\
P2  & D7  & Lap          & 2.2  & 100.0 & 2.0 & 0   & 2.0 \\
P3  & D5  & Laptop stand & 8.9  &  93.3 & 6.5 & 0.8 & 6.8 \\
P4  & D1  & Wood         & 6.7  & 100.0 & 5.0 & 0   & 5.3 \\
P5  & D2  & Plastic      & 5.2  & 100.0 & 4.1 & 0   & 4.2 \\
P6  & D6  & Lap          & 7.4  & 100.0 & 5.7 & 0.8 & 5.9 \\
P7  & D4  & Mouse pad    & 5.9  &  93.3 & 4.7 & 0.7 & 4.9 \\
\cmidrule(lr){1-8}
\multicolumn{3}{@{}l}{\textbf{Avg.}} & 6.2 & 98.1 & 4.8 & 0.4 & 5.0 \\
\midrule
\multicolumn{8}{@{}l}{\emph{End-to-end}} \\
P8  & D8  & Caf\'{e} table & 4.4  & 93.3 & 2.1 & 1.4 & 2.5 \\
P9  & D9  & Lap            & 15.6 & 80.0 & 10.0 & 11.5 & 10.9 \\
P10 & D10 & Mattress       & 9.6  & 86.7 & 4.3 & 2.2 & 5.0 \\
\cmidrule(lr){1-8}
\multicolumn{3}{@{}l}{\textbf{Avg.}} & 9.9 & 86.7 & 5.3 & 4.8 & 6.0 \\
\bottomrule
\end{tabular*}
\end{table}

\noindent \textbf{Failure analysis: P9.}
P9 exhibits the highest raw CER among all ten participants: 15.6\% for passwords, 10.0\% for sentences, and 10.9\% overall. It is also the only case in which language-model augmentation degrades sentence recovery, increasing CER from 10.0\% to 11.5\%. We observe that P9's shorter inter-keystroke intervals can cause three or four consecutive 300\,ms classification windows to overlap, contaminating each target window with multiple neighboring keystrokes. The resulting substitutions may span distant keyboard positions and form valid English words that the language model has little basis to reject. In one representative sentence, five of the seven raw character errors formed valid words; although the language model corrected two non-word errors, it increased the total error count from seven to nine. This pattern occurs in three of P9's ten test sentences and accounts for the higher CER after language-model rescoring.

\subsection{Label-Free User Profiling}
\label{sec:brutus-user}

In shared-console settings, a recurring group may take turns at a logged-in MacBook while its account and UID remain fixed. From the IMU stream, \attack estimates the number of user profiles and assigns subsequent typing segments to them.

\noindent \textbf{Attack preparation.}
Eight participants (P1--P8; four men and four women, ages 22--35) type the Section~\ref{sec:char-setup} passage on D1 on a wood desk. After filtering backspace, return, punctuation, and modifier combinations, each contributes 1{,}188 windows: the first 891 form nine discovery segments and the remaining 297 form three subsequent segments, totaling 72 and 24 segments. Fixing text, device, and surface isolates typing dynamics.

\noindent \textbf{Segment representation.}
Idle gaps divide the onset stream into typing segments of approximately 99 windows. For each window, \attack computes all 15 pairwise Pearson correlations among the six IMU channels; seven distribution summaries per pair yield an $\ell_2$-normalized, 105-dimensional segment representation. Inter-axis correlations characterize coupled multiaxis motion~\cite{bao2004activity} and are invariant to the amplitude scaling caused by strike force (Section~\ref{sec:char-typing-force}), while retaining relative propagation across axes. Aggregation captures recurring hand posture, finger assignment, and striking motion~\cite{monrose2000keystroke,gascon2014continuous,giuffrida2014sensed}. Clustering and attribution both use the full 105-dimensional space.

\noindent \textbf{Profile discovery and attribution.}
With the representation and criterion fixed before evaluation, \attack applies Lloyd's $k$-means~\cite{lloyd1982} and uses prediction strength to test reproducibility across data splits~\cite{tibshirani2005prediction}, selecting the largest $k$ with $\overline{\mathrm{PS}}(k)\geq0.80$. For adjacent accepted solutions, \emph{nested purity}---standard purity relative to the coarser partition~\cite{manning2008introduction}---equals 1.0 when every finer cluster lies within one coarser cluster. \attack then refits on all discovery segments and assigns subsequent segments to the nearest centroid in the full space.

\begin{figure}[t]
  \centering
  \includegraphics[width=\columnwidth]{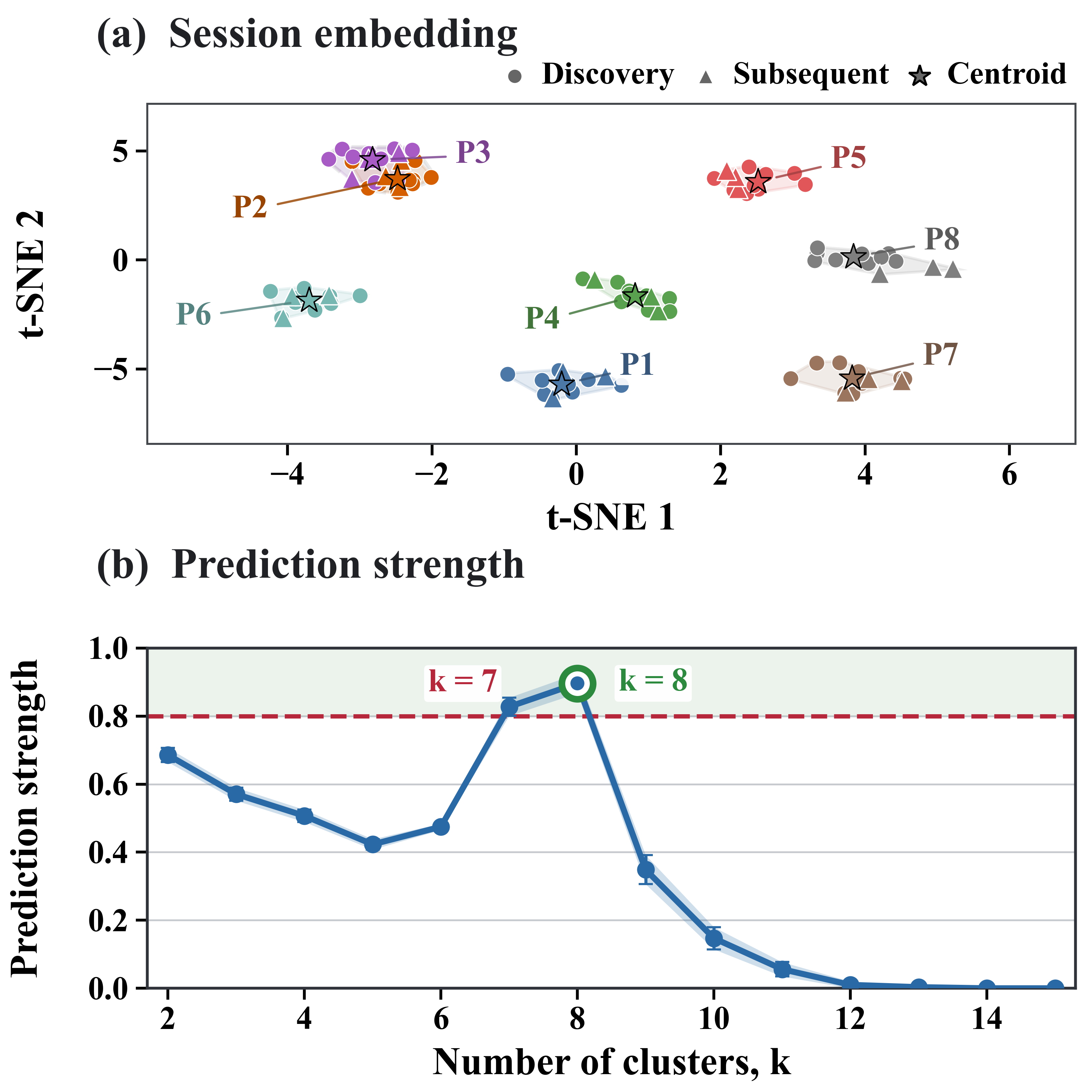}
  \caption{Label-free user profiling. (a)~t-SNE of discovery segments (circles), subsequent segments (triangles), and centroids (stars); colors show participants after evaluation alignment. (b)~Prediction strength versus $k$; the dashed line marks the 0.80 threshold and the ring marks $k=8$.}
  \label{fig:user-cm}
\end{figure}

\noindent \textbf{Attack output.}
Figure~\ref{fig:user-cm}(b) shows $\overline{\mathrm{PS}}(7)=0.827$ and $\overline{\mathrm{PS}}(8)=0.896$, both above the selection threshold; the score falls to 0.349 at $k=9$. \attack therefore selects the largest accepted solution, $\hat{k}=8$. The transition from $k=7$ to $k=8$ has nested purity 1.0: the joint P2/P3 cluster divides into separate profiles while the other six clusters remain intact. After Hungarian matching aligns the discovered clusters with participant identifiers for evaluation~\cite{kuhn1955hungarian}, the eight clusters correspond to P1--P8, and all 72 discovery segments and 24 subsequent segments are assigned to the correct profiles.

Figure~\ref{fig:user-cm}(a) visualizes this hierarchy using t-SNE~\cite{maaten2008tsne}. P2 and P3 form the closest pair. Both are women aged 24 and 25 with similar builds and type with extended fingers held relatively flat against the keys. This shared posture couples impacts into the chassis similarly, merging their profiles at $k=7$. They use different fingers for some keys, however, changing impact direction and cross-axis propagation. Aggregating approximately 99 keystrokes per segment exposes these persistent differences, separating the profiles at $k=8$ and assigning subsequent segments accordingly. Thus, $k=7$ captures their shared coarse-grained posture, whereas $k=8$ distinguishes the user-specific mechanics introduced by their different finger assignments.

\subsection{Label-Free Environment Profiling}
\label{sec:brutus-env}

Because the support condition changes keystroke propagation across IMU axes, \attack groups recurring typing segments into environment profiles and attributes subsequent segments.

\noindent \textbf{Attack preparation.}
P1 types the Section~\ref{sec:char-setup} passage on D1 across nine surfaces: wood, steel, plastic, glass, mouse pad, laptop stand, leather sofa, soft mattress, and lap. After filtering, each environment contributes 1{,}188 windows: 891 form nine discovery segments and 297 form three subsequent segments, totaling 81 and 27 segments. Fixing typist, device, and content isolates the support condition.

\noindent \textbf{Profile discovery and attribution.}
\attack reuses the segment representation and prediction-strength procedure from Section~\ref{sec:brutus-user}, assigning subsequent segments to discovery centroids. As Section~\ref{sec:char-desk} shows, stiffness, damping, and contact geometry shape propagation among IMU channels, producing recurring environment profiles.

\begin{figure}[t]
  \centering
  \includegraphics[width=\columnwidth]{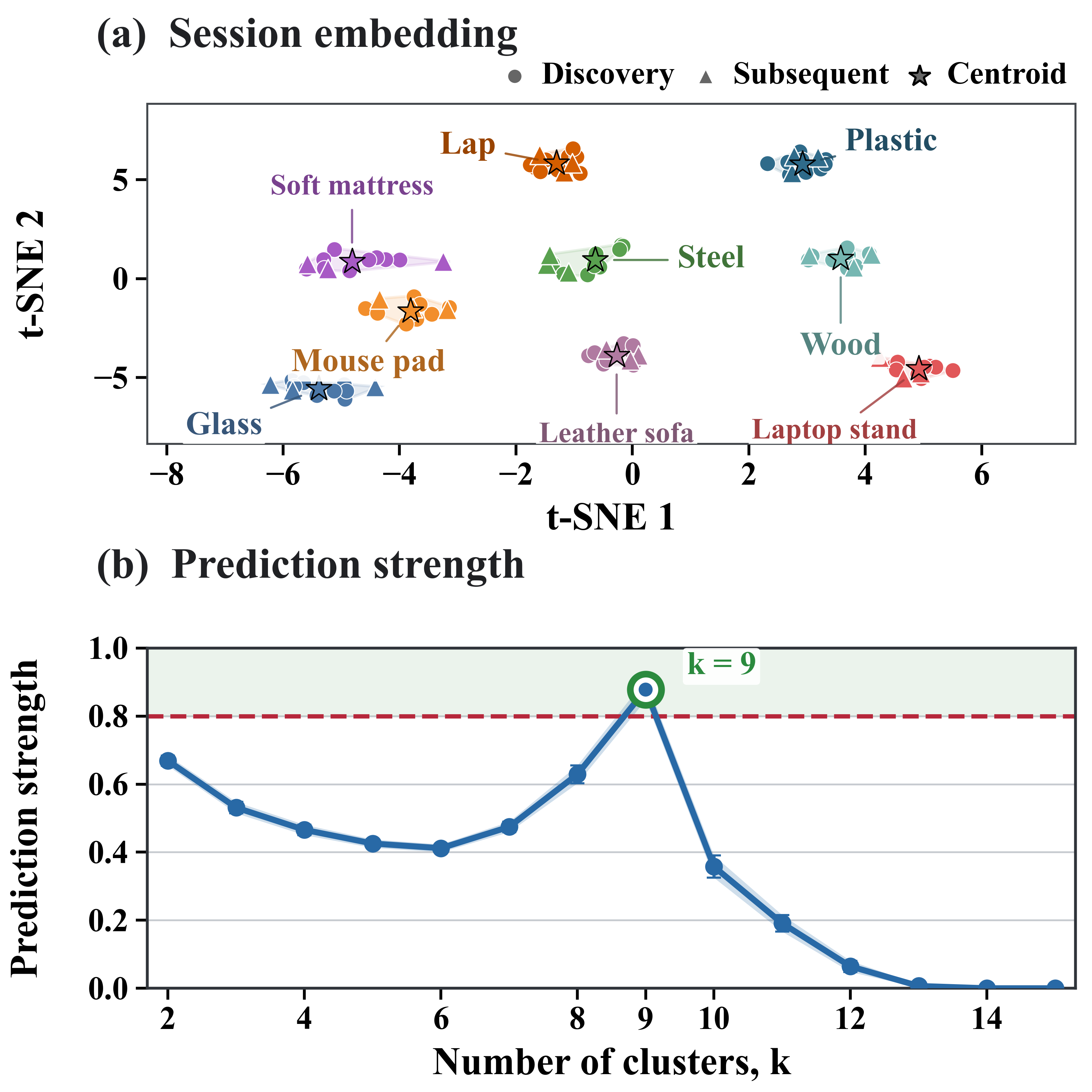}
  \caption{Label-free environment profiling. (a)~t-SNE of discovery segments (circles), subsequent segments (triangles), and centroids (stars); colors show environments after alignment. (b)~Prediction strength versus $k$; the dashed line marks 0.80 and the ring marks $k=9$.}
  \label{fig:surface-cm}
\end{figure}

\noindent \textbf{Attack output.}
Figure~\ref{fig:surface-cm}(b) gives prediction strength 0.878 at $k=9$, so \attack selects $\hat{k}=9$. After Hungarian matching, the nine clusters correspond to the nine environments, and all 81 discovery and 27 subsequent segments are assigned correctly.

In Figure~\ref{fig:surface-cm}(a), each surface forms a distinct discovery region and subsequent segments lie near the corresponding centroid, showing persistent support-dependent propagation.

%% file: sections/sec_discussion.tex
\section{Discussion}
\label{sec:discussion}

\subsection{Other Exploitations}
\label{sec:disc-other}

\noindent \textbf{Trackpad side channel.}~
The built-in IMU can capture vibrations from Force Touch taps, clicks, and swipes~\cite{marquardt11,miluzzo12}; whether they reveal click targets or gestures remains future work.


\noindent \textbf{Other devices.}~
Similar leakage may arise where a keyboard and software-readable IMU share a rigid structure; for example, iOS \texttt{CMMotionManager} exposes motion data at approximately 100\,Hz in the foreground~\cite{apple-coremotion}. Cross-platform evaluation remains future work.

\subsection{Limitations}
\label{sec:disc-limitations}

The IMU path requires the process to run under the active console user's UID and that account to belong to the administrator group (Section~\ref{sec:char-access}); the initial, typically sole Mac user commonly satisfies both. GUI processes, SSH sessions under that UID, and same-UID \texttt{LaunchAgent}s can acquire the full 800\,Hz stream.

A secondary SSH or \texttt{su} user, even an administrator, receives \texttt{kIOReturnNotPrivileged} at \texttt{IOHIDDeviceOpen} because its UID differs from the console owner. Fast User Switching also disconnects the stream within three seconds, without an error or handle invalidation.



\subsection{Mitigation}
\label{sec:disc-mitigation}

We propose three complementary system- and driver-level defenses.

\noindent \textbf{Access-control enforcement.}~
The \texttt{AppleSPUHIDDevice} path lacks the mediation applied to comparable mobile sensor interfaces. Extending TCC to this IOKit path, analogous to protecting Intel power interfaces~\cite{lipp2021platypus}, would block unauthorized reads. Equivalent checks on \texttt{HIDIdleTime} and \texttt{CGEventSourceSecondsSinceLastEventType} would also remove the content-free timing oracle used to synchronize keystrokes.

\noindent \textbf{Rate limiting.}~
Sensor rate limiting has emerged as a standard defense against motion-based side channels across major mobile platforms: iOS restricts \texttt{CMMotionManager} to approximately 100\,Hz~\cite{apple-coremotion}, while Android~12 mandates the \texttt{HIGH\_SAMPLING\_RATE\_SENSORS} permission for sampling rates exceeding 200\,Hz~\cite{android-sensors}.

Following Section~\ref{sec:char-setup}, P1 types on D1 on a wood desk at 800, 200, 100, and 50\,Hz. Every rate uses the same 1{,}842-character corpus: 50 passwords each of lengths~8 and~9 plus 12 sentences. Training uses all length-8 passwords, 30 length-9 passwords, and 9 sentences (1{,}406 characters); testing uses the remaining 20 length-9 passwords and 3 sentences (436 characters). We train from scratch and test at the same rate using the five-fold session-level protocol from Section~\ref{sec:brutus-overview}.

\begin{table}[t]
\centering
\caption{Keystroke recovery under rate limiting (\%), averaged over five same-rate session-level folds.}
\label{tab:rate-limit}
\renewcommand{\arraystretch}{1.15}
\setlength{\tabcolsep}{3pt}
\footnotesize
\begin{tabular*}{\columnwidth}{@{\extracolsep{\fill}}l cc cc@{}}
\toprule
& \multicolumn{2}{c}{\textbf{Password}} & \multicolumn{2}{c}{\textbf{Sentence}} \\
\cmidrule(lr){2-3} \cmidrule(lr){4-5}
\textbf{Rate} & \textbf{CER} & \textbf{Top-5} & \textbf{CER} & \textbf{+LM} \\
\midrule
800\,Hz & 3.33  & 96.00 & 4.14  & 0.86  \\
200\,Hz & 26.56 & 31.00 & 23.67 & 18.59 \\
100\,Hz & 36.38 & 8.00  & 25.38 & 16.78 \\
50\,Hz  & 62.00 & 0.00  & 43.36 & 35.86 \\
\bottomrule
\end{tabular*}
\end{table}

Table~\ref{tab:rate-limit} shows substantial degradation: password Top-5 recovery falls from 96.00\% at 800\,Hz to 31.00\%, 8.00\%, and 0\% at 200, 100, and 50\,Hz, while password CER reaches 62.00\% at 50\,Hz. Language-model-assisted sentence CER rises from 0.86\% to 35.86\%. Rate limiting therefore suppresses fine-grained vibration information and complements access control.

\noindent \textbf{Noise injection.}~
Calibrated driver-level noise requires no application cooperation. Section~\ref{sec:char-noise} shows that additive, chassis-coupled vibration impairs cross-device discrimination; targeting the keystroke band could obscure this channel while preserving coarse orientation and motion sensing.

%% file: sections/sec_related.tex
\section{Related Work}
\label{sec:related}

Prior inference techniques span three targets: user identity, device environment, and keystroke content. We organize related work accordingly and compare sensor placement, external hardware, active probing, label requirements, and user enrollment.

\subsection{User Identification}

\noindent \textbf{Keystroke dynamics.} Classical keystroke biometrics construct a user template from key-hold and inter-key timings, and then verify a claimed identity or identify one typist from an enrolled gallery~\cite{monrose2000keystroke,gunetti2005freetext,killourhy2009comparing}. TypeNet and Type2Branch learn embeddings that transfer to identities excluded from model training~\cite{acien2022typenet,gonzalez2025type2branch}. Nevertheless, both classical templates and learned embeddings require identity labels during training or enrollment: every test user must first contribute an identity-linked reference set. This per-user registration makes the attack expensive to scale and leaves an attacker who obtains only unlabeled sessions without the references needed to separate those sessions by user.

\noindent \textbf{Touch and motion behavior.} Touchalytics and SilentSense authenticate a phone owner from touchscreen gestures and touch-induced device motion~\cite{frank2013touchalytics,bo2013silentsense}. Other systems combine touch or keystroke events with a phone's accelerometer and gyroscope~\cite{gascon2014continuous,giuffrida2014sensed,sitova2016hmog}, while BehaveFormer learns a supervised representation from keystroke and IMU sequences~\cite{senarath2023behaveformer}. Owner-verification systems answer whether the current operator matches one enrolled account; gallery-based systems select among identities registered in advance. In either case, the attacker must know the candidate identities and collect labeled reference data for each new target, so the method cannot discover the number or membership of users directly from unlabeled observations.

\noindent \textbf{Wearables and external sensing.} WACA records typing motion from a smartwatch worn by the user, and Lee et al. actively vibrate a smartwatch and measure the wearer's response~\cite{acar2018waca,lee2021smartwatchauth}. Roth et al. place a microphone near the keyboard to identify typists from keystroke sounds~\cite{roth2015keystrokesound}; VibWrite attaches an actuator and vibration sensors to the input surface~\cite{lu2017vibwrite}. These prerequisites sharply restrict deployability: an attacker must place a sensor near the victim, control a device worn by the victim, or touch and instrument the input surface. None can be launched solely by local malware already running on an otherwise unmodified target computer.

\noindent \textbf{Speech-induced motion.} Gyrophone recognizes a fixed set of speakers by measuring gyroscope responses to speech from an external loudspeaker~\cite{michalevsky2014gyrophone}; Spearphone and AccelEve exploit coupling between a phone's own loudspeaker and accelerometer~\cite{anand2021spearphone,ba2020acceleve}. These attacks require speech playback or speaking, sufficiently strong loudspeaker--sensor coupling, and labeled samples for a predetermined set of talkers. Changing that set requires another identity-linked collection and enrollment process.

\noindent \textbf{Comparison with \attack.} Existing user-identification methods generally rely on identity labels and per-user enrollment. Methods using a smartwatch, an external microphone, or active vibration further require the attacker to control a victim-worn device, place a sensor near the victim, or instrument the input surface, making them difficult to deploy through local malware alone on the target computer. \attack uses only the MacBook's built-in IMU to estimate the number of user clusters from unlabeled sessions and group sessions by user, without identity labels, per-user enrollment, or external sensing hardware.

\subsection{Environment Identification}

\noindent \textbf{Active vibration and acoustics.} VibePhone and Diaconita et al. activate a phone's vibration motor and classify the accelerometer response~\cite{cho2016vibephone,diaconita2015position}; GripSense combines touch and gyroscope signals for hand-posture sensing and pulses the vibration motor for pressure inference~\cite{goel2012gripsense}. Hasegawa et al. emit tones through a phone's built-in speaker and analyze the signal returned to its microphone~\cite{hasegawa2017placement}. Kunze and Lukowicz combine vibration, acceleration, sound, and a downward-facing extra loudspeaker~\cite{kunze2007symbolic}. These active paths require the attacker to control an actuator and inject a known probe before measurement, preventing passive inference from ordinary device use. Their measurements are tied to the probe hardware, orientation, and contact geometry, while expanding the target class set requires another labeled collection campaign.

\noindent \textbf{Specialized material sensors.} Strese et al. use a custom haptic stylus with accelerometers, microphones, force sensors, external acquisition hardware, and close-up surface images~\cite{strese2017multimodal}. Harrison and Hudson build a separate multispectral optical prototype that actively illuminates nearby material~\cite{harrison2008material}. Such instrumentation must be installed on or brought into contact with the target environment, making the attack conspicuous and dependent on prior physical access; it cannot be executed by software on an unmodified victim computer. Both approaches also require labeled examples for their predefined material classes.

\noindent \textbf{Comparison with \attack.} Existing environment-identification methods either control a vibrator or loudspeaker to emit a probe, or use custom haptic and optical instruments, and train on labeled examples of predefined material, location, or posture classes. These prerequisites require the attacker to control an actuator or physically access the environment to deploy specialized instrumentation, limiting stealth and deployability. \attack uses only the MacBook's built-in IMU to estimate the number of environment clusters from unlabeled sessions and group sessions by physical environment during natural typing, without environment labels, external instrumentation, or active sound or vibration.

\subsection{Keystroke Recovery}

Keystroke eavesdropping attacks can be divided into physical side-channel attacks (PSCAs) and software side-channel attacks (SSCAs).

\noindent \textbf{Acoustic, electromagnetic, and wireless PSCAs.} Acoustic attacks record key-dependent sound using a nearby microphone or an accessible audio stream~\cite{asonov04,zhuang05,slater19,tu23}. Their models are sensitive to the keyboard, typist, microphone placement, and ambient noise; without access to a suitable audio path, the leakage is unavailable. Electromagnetic and wireless attacks instead capture keyboard emanations or key-induced perturbations to Wi-Fi and radio signals~\cite{vuagnoux09,ali15,fang23,radkey26}. They require an electromagnetic receiver, controllable wireless infrastructure, or a tag and reader near the victim, exposing the attack to physical discovery and preventing deployment by target-local software alone.

\noindent \textbf{Mechanical-vibration PSCAs.} A smartphone placed on the same desk can sense keyboard vibration~\cite{marquardt11}; smartwatches and headphones can capture related wrist or head motion~\cite{liu15,maiti16,overhear23}. The sensing device must lie on the vibration path between the keyboard and the attacker. Moving the phone, changing the support material, or removing the wearable breaks that path, while pre-positioning the hardware requires physical proximity or control of a victim-owned accessory.

\noindent \textbf{Built-in sensors.} Camera and microphone streams exposed to conferencing applications support video- and audio-based recovery~\cite{compagno17,yang23video}, but macOS places camera and microphone access behind Transparency, Consent, and Control authorization~\cite{apple-tcc}. On phones and watches, motion attacks recover keys, PINs, and longer text through public sensor APIs~\cite{cai11,xu12taplogger,owusu12,miluzzo12,mehrnezhad17,ping15textlogger}. That deployment path does not carry over to MacBooks: Apple's supported Core Motion interface is unavailable on macOS~\cite{apple-coremotion}, leaving no documented third-party API for the chassis IMU.

\noindent \textbf{SSCAs.} Network-based attacks infer input from encrypted-traffic timing and size patterns, as in search-engine autocomplete~\cite{monaco19}; they observe only inputs that trigger distinguishable communication, so offline, buffered, or purely local typing leaves no corresponding trace, and protocol or application changes can invalidate the model. Timing attacks exploit inter-keystroke intervals in interactive sessions~\cite{song01,qiu26keytar}. Because an interval does not directly identify either key, reconstruction depends on candidate corpora, dictionaries, or language priors and degrades on random secrets and atypical timing. Cache attacks locate keystroke-related execution paths and data accesses~\cite{gruss15,schwarzl23lbta}; they require processor-specific cache primitives and stable application binaries, and software updates or architectural changes can invalidate the learned templates.

\noindent \textbf{Comparison with \attack.} On MacBooks, existing physical attacks either require sensors to be placed near the victim or require camera or microphone permission; inertial attacks on mobile devices rely on public motion-sensor APIs, but macOS provides no corresponding public interface for reading the MacBook's built-in IMU. Among software attacks, traffic-based methods apply only to online interactive settings in which each input triggers network transmission, leaving no per-keystroke traffic for local, offline, or buffered input; timing methods require key-labeled samples from the target user to train or adapt a personal timing model; cache attacks that directly reveal key values must first locate keycode-related code or data addresses in the target program and depend on a specific application, binary version, shared mapping, and processor cache primitives, so application updates or platform changes invalidate the template. \attack reads the factory-installed IMU through an undocumented IOKit path and recovers keys from keystroke-induced vibrations independent of the target application, without labeled training data from the target user, external hardware, active sound or vibration, or camera or microphone permission. The same data stream also supports label-free user and environment analysis.

%% file: sections/sec_conclusion.tex
\section{Conclusion}
\label{sec:conclusion}

In this paper, we present \attack, an unprivileged side-channel attack framework that exploits the undocumented IMU built into Apple MacBooks. \attack reads raw IMU data through IOKit without root privileges or runtime privilege elevation and combines the sensor stream with two content-free system metadata interfaces to infer keystrokes, typists, and laptop placement surfaces. Against three held-out participants on unseen devices, \attack achieves a character-level accuracy of 89.1\% to 97.5\%. Without user or environment labels, it also discovers user and environment profiles and assigns subsequent segments to their corresponding profiles. Our results show that platform sensor access controls must extend to undocumented, vendor-internal sensor interfaces.

%% file: sections/sec_ethics.tex
\section*{Ethical Considerations}
\label{app:ethics}

\noindent\textbf{Participants and data handling.}
Our evaluation involved ten adult participants in controlled experiments. All
participants provided informed consent before data collection and received
US\$100 for their participation. They typed passages supplied for the study,
randomly generated alphanumeric passwords, and test sentences. We refer to
participants and devices only by the identifiers P1--P10 and D1--D10. The
research team stores the source data, and participants may request deletion of
their records. Before release, the raw IMU data from
Section~\ref{sec:characterization} are de-identified by removing names, account
information, device serial numbers, absolute file paths, original collection
times, and other identifying metadata. The participant data, trained models,
and end-to-end attack implementation from Section~\ref{sec:brutus} are retained
within the research team.

\noindent\textbf{Potential impact and risk management.}
The stakeholders in this work include the study participants, users of
IMU-equipped MacBooks, Apple, and the security research community. All
experiments were conducted with consenting participants on controlled devices
using inputs supplied for the study. The IMU access path has dual-use
implications: it enables researchers to evaluate the platform's sensor access
controls, but it could also allow malicious local software to infer typed
content, typist characteristics, and the laptop's physical context. The public
artifact contains the access-control audit and physical-leakage
characterization from Section~4. These materials support independent
validation of the underlying system and physical evidence. The participant
data, trained models, and complete attack pipeline from Section~5 remain
internal.

\noindent\textbf{Responsible disclosure and publication.}
We disclosed the IMU access-control issue and the resulting side-channel
vectors to Apple in March 2026. In May 2026, Apple acknowledged the report and
confirmed that it had reproduced the IMU data leakage. At the time of
submission, Apple was investigating the root cause and developing mitigations.
Publishing these findings enables independent examination of the exposed
sensor path and provides evidence for platform-level access controls, sampling
rate limits, and user-visible permission mechanisms.

%% file: sections/sec_open_science.tex
\section*{Open Science}
\label{app:open-science}

An anonymized artifact is available at
\url{https://anonymous.4open.science/#!/r/submission-artifact-604F/}. It includes a
demonstration video recorded by the authors and the complete experimental
materials for the IMU characterization in Section~\ref{sec:characterization}.
The materials cover the access-control audit, key-position and typing-force
characterization, supporting-surface characterization, cross-device
evaluation, and noise evaluation. For each experiment, the artifact provides
de-identified raw six-axis IMU recordings, the experimental protocol and
configuration, labels and provenance metadata, analysis scripts, and expected
outputs. The accompanying documentation maps these materials to the
corresponding figures, tables, and findings in the paper.

The Section~4 artifact supports independent verification of the access-control
and physical-leakage properties underlying the attacks in
Section~\ref{sec:brutus}. Reacquiring the sensor stream and repeating the
access-control audit require a compatible Apple Silicon MacBook with a built-in
IMU and an appropriate macOS configuration. The packaged recordings and
analysis scripts support offline reproduction of the Section~4 analyses. The
public release covers Section~4. The participant datasets, trained models, and
end-to-end attack implementation from Section~5 are retained by the research
team. All released files, including the demonstration video, are anonymized to
remove participant, device, institution, and author identifiers.